\documentclass[modern]{aastex63}
\usepackage{graphics,amsmath,amsbsy}
\usepackage{enumitem}
\setlist[enumerate]{nosep}
\setlist[itemize]{topsep=0pt}

\newcommand\eps{\epsilon}
\newcommand{\lapprox} {\, \lower3pt\hbox{$\sim$}\llap{\raise2pt\hbox{$<$}}\,}
\newcommand{\gapprox} {\, \lower3pt\hbox{$\sim$}\llap{\raise2pt\hbox{$>$}}\,}

\definecolor{mrk}{RGB}{160,0,0}

\newcommand\mmatrix[1]{\textsf{\textbf{#1}}}
\renewcommand{\vec}[1]{ \protect {{\mathbf{\boldsymbol{#1}}}}}

\shorttitle{Sub-second time evolution of Type III bursts}
\shortauthors{Chen, Kontar, Chrysaphi et al.}
\begin{document}
\title{Sub-second time evolution of Type III solar radio burst sources at fundamental and harmonic frequencies}

\author[0000-0002-1810-6706]{Xingyao Chen}
\affil{Key Laboratory of Solar Activity, National Astronomical Observatories Chinese Academy of Sciences, Beijing 100101, China}
\affil{School of Physics \& Astronomy, University of Glasgow, Glasgow, G12 8QQ, UK}

\author[0000-0002-8078-0902]{Eduard P. Kontar}
\affil{School of Physics \& Astronomy, University of Glasgow, Glasgow, G12 8QQ, UK}

\author[0000-0002-4389-5540]{Nicolina Chrysaphi}
\affil{School of Physics \& Astronomy, University of Glasgow, Glasgow, G12 8QQ, UK}

\author[0000-0001-6583-1989]{Natasha L.S. Jeffrey}
\affil{Department of Mathematics, Physics and Electrical Engineering, Northumbria University, Newcastle upon Tyne, NE1 8ST, UK}

\author[0000-0003-2291-4922]{Mykola Gordovskyy}
\affiliation{School of Physics \& Astronomy, University of Manchester, Manchester M13 9PL, UK}

\author{Yihua Yan}
\affil{Key Laboratory of Solar Activity, National Astronomical Observatories Chinese Academy of Sciences, Beijing 100101, China}
\affil{School of Astronomy and Space Sciences, University of Chinese Academy of Sciences, Beijing 100049, China}

\author[0000-0003-2047-9664]{Baolin Tan}
\affil{Key Laboratory of Solar Activity, National Astronomical Observatories Chinese Academy of Sciences, Beijing 100101, China}
\affil{School of Astronomy and Space Sciences, University of Chinese Academy of Sciences, Beijing 100049, China}

\begin{abstract}
Recent developments in astronomical radio telescopes opened new opportunities in imaging and spectroscopy of solar radio bursts at sub-second timescales.
Imaging in narrow frequency bands has revealed temporal variations in the positions and source sizes that do not fit into the standard picture of type III solar radio bursts, and require a better understanding of radio-wave transport.
In this paper, we utilise 3D Monte Carlo ray-tracing simulations that account for the anisotropic density turbulence in the inhomogeneous solar corona to quantitatively explain the image dynamics at the fundamental (near plasma frequency) and harmonic (double) plasma emissions observed at $\sim$32~MHz.
Comparing the simulations with observations, we find that anisotropic scattering from an instantaneous emission point source can account for the observed time profiles, centroid locations, and source sizes of the fundamental component of type III radio bursts (generated where $f_{pe} \approx 32$~MHz).
The best agreement with observations is achieved when the ratio of the perpendicular to the parallel component of the wave vector of anisotropic density turbulence is around 0.25. Harmonic emission sources observed at the same frequency ($\sim$32~MHz, but generated where $f_{pe} \approx 16$~MHz) have apparent sizes comparable to those produced by the fundamental emission, but demonstrate a much slower temporal evolution. The simulations of radio-wave propagation make it possible to quantitatively explain the variations of apparent source sizes and positions at sub-second time-scales both for the fundamental and harmonic emissions, and can be used as a diagnostic tool for the plasma turbulence in the upper corona.
\end{abstract}

\section{Introduction}
\label{sec01}
Solar radio bursts are commonly considered to be a signature of acceleration and propagation of non-thermal electrons in the solar corona \citep[e.g.][]{1958SvA.....2..653G, 1985ARA&A..23..169D}. 
In the standard type III solar radio burst scenario, non-thermal electrons propagate away from the Sun and generate Langmuir waves\citep[e.g.][]{1958SvA.....2..653G,1983SoPh...89..403G,2016PhRvE..93c3203Y}, so that the radio emission is progressively
produced at lower frequencies as the electrons responsible for radio emission propagate away from the Sun. The radio emission is produced at the fundamental and harmonic (twice the local plasma frequency) frequencies, so observations at the same frequency examine the harmonic emission from distances further away from the Sun.
At the same time, various propagation effects -- including the refraction due to plasma density gradients and scattering by small-scale density fluctuations -- significantly affect the apparent properties of the radio sources, including their time evolution, position, and size \citep{1971A&A....10..362S, 1999A&A...351.1165A, 2017NatCo...8.1515K}. Scattering in the inhomogeneous solar corona is also considered as a possible explanation for the decrease in the apparent brightness temperature of the quiet Sun at about 30 MHz \citep{1971A&A....12..435A,2007ApJ...671..894T}. 
Therefore, radio-wave scattering needs to be taken into account in the analyses of solar radio observations.

Scattering of radio waves in the solar corona has been extensively studied since the first solar radio observations \citep{1965BAN....18..111F, 1971A&A....10..362S, 1972A&A....18..382S, 1974SoPh...35..153R, 1977A&A....61..777B, 1983PASAu...5..208R, 1994ApJ...426..774B, 1999A&A...351.1165A, 2007ApJ...671..894T,2020ApJ...889L..25R}. Consequently, \cite{1965BAN....18..111F} performed the first numerical study of scattering on small-scale density inhomogeneities to estimate the apparent sizes and locations of solar radio-sources. By accounting for absorption, irregular refraction by large-scale inhomogeneities,
and isotropic scattering, \cite{1971A&A....10..362S},  \cite{1972A&A....18..382S}, and \cite{1974SoPh...35..153R} extended the ray-tracing method to study the arrival time, intensity, and angular broadening of the scattered images.
\cite{1999A&A...351.1165A} considered the propagation of radio waves in an anisotropic, statistically inhomogeneous plasma using geometrical optics and the Hamilton equations, and successfully reproduced many features of radio bursts. Later, \cite{2007ApJ...671..894T} developed Monte Carlo simulations to study isotropic scattering effects focusing on the directivity of interplanetary type III bursts at frequencies ranging between $100-200$~kHz. \cite{2018ApJ...857...82K} simulated the scattering effects on low-frequency plasma radio emission and demonstrated that type III bursts can be used as a diagnostic tool for plasma density variations in the solar wind.

Recent observations by \cite{2018ApJ...857...82K} suggest that radio-wave scattering effects in the solar corona and interplanetary space dominate the observed duration of fundamental radio sources.  Most of the radio-wave scattering simulations assumed that the density inhomogeneities are isotropic, but \cite{2019ApJ...884..122K} found that the time duration and source size cannot be simultaneously reproduced using isotropic density fluctuations, and suggested that radio-wave scattering is strongly anisotropic.

\cite{2017NatCo...8.1515K} and \cite{2018SoPh..293..115S} observed the centroid positions, sizes, and areal extents of the fundamental and harmonic sources of a Type III-IIIb burst, and demonstrated that scattering effects dominate the observed spatial characteristics of radio burst images.  \cite{2018ApJ...868...79C} illustrated that the spatial separation observed between the sources of split-band Type II bursts is consistent with radio-wave scattering effects.

For the first time, in this study, Monte Carlo simulations of radio-wave propagation are used to investigate the time evolution, positions, and sizes of the apparent solar radio burst sources at sub-second scales. The simulations are performed both for the fundamental and harmonic components, with the aim of explaining the properties of type III-IIIb solar radio bursts observed by the LOw-Frequency ARray \citep[LOFAR;][]{2013A&A...556A...2V}.

The article is arranged as follows. Section \ref{sec02} describes the anisotropic scattering and ray-tracing model. Section \ref{sec03} describes the simulation results and their comparison with the radio burst observations. Section \ref{sec04} presents the discussion and conclusions.

\section{Radio-Wave Scattering Simulations} \label{sec02}
\subsection{Equations} \label{sec0201}
Radio waves with angular frequency $\omega$ and wavevector $k$ 
in an unmagnetised plasma follow the dispersion relation $\omega ^2= \omega_{pe}^2 +k^2c^2$, where $\omega_{pe}=\sqrt{4\pi e^2 n (\vec{r})/m_e}$ is the electron plasma frequency.  
Both the fundamental emission ($\omega \sim \omega_{pe}$) and second-harmonic radio waves ($\omega \sim 2\omega_{pe}$, produced farther away from the Sun)
have their frequencies close to the local plasma frequency $\omega_{pe}$, and are hence
sensitive to the electron density fluctuations via the plasma refractive index $\sqrt{1-\omega _{pe}^2/\omega^2}$ \citep[see, e.g.][]{2012wop..book.....P}.

In this study we use the same numerical approach as in \cite{2019ApJ...884..122K}. The density fluctuations in the solar corona are assumed to be axially symmetric with respect to the local radial direction (the direction of the guiding magnetic field), so that the spectrum can be parameterized as a spheroid in wavevector-space:
\begin{equation}\label{eq:S_ani}
S(\vec q) = S \, \left( \left [ {q_{\perp}}^2+{\alpha}^{-2}{q_\parallel}^2 \right ]^{1/2}\right) \,\,\, ,
\end{equation}
where $\alpha =h_\perp/h_\parallel$ is the ratio of perpendicular and parallel correlation lengths, leading to the diffusion tensor components $D_{ij}$:
\begin{equation}
\label{eq:d_ij_aniso_k}
D_{ij} = \left[\frac{{A}^{-2}_{ij}}{(\vec k \mmatrix{A}^{-2} \vec k)^{1/2}}-\frac{(\mmatrix{A}^{-2}\vec k)_i(\mmatrix{A}^{-2}\vec k)_j}{(\vec k \mmatrix{A}^{-2} \vec k)^{3/2}}\right]
\frac { { \omega } _ { p e } ^ { 4 } } 
{ 32\pi { \omega } c^2 } \alpha \int_0^{\infty} {\widetilde q}^3 \, S(\widetilde q) \, {d\widetilde q} \,\,\, .
\end{equation}
Here $\mmatrix{A}$ is the anisotropy matrix
\begin{equation} \label{eq:A_matrix}
\mmatrix{A}=
{\left( \begin{array}{ccc}
1 & 0 & 0 \\
0 & 1 & 0 \\
0 & 0 & {\alpha}^{-1} \\
\end{array} \right )},
\end{equation}
and $\vec{\widetilde q}= \mmatrix{A}\vec q$ (so that $\vec q= \mmatrix{A}^{-1} \vec{\widetilde q}$).
The scattering tensor (Equation \ref{eq:d_ij_aniso_k}) depends on the spectrum-averaged wavenumber of the density fluctuations
\begin{equation}\label{eq:q_bar}
\overline{q\eps ^2}=\overline{q\delta n ^2}\frac{1}{n^2} =\frac{1}{n^2}\int_0^{\infty}  {\widetilde q}^3 \, S(\widetilde q) \, \frac{4\pi d\widetilde q}{(2\pi)^3}
\end{equation}
and the anisotropy parameter $\alpha$. Therefore, knowing the mean wavenumber $\overline{q\eps^2}$, one can describe the scattering of radio waves in a turbulent solar plasma \citep{2019ApJ...884..122K}.

In-situ \citep{1983A&A...126..293C,2020ApJS..246...57K} and remote-sensing radio observations \citep{2001SSRv...97....9W} suggest that the density fluctuations in the interplanetary space have an inverse power-law spectrum $S(q)\propto q^{-(p+2)}$, with the spectral index $p$ close to $5/3$. The power-law is normally observed over a broad inertial range, from the outer scales $l_0 = 2\pi/q_0$ to the inner scales $l_i=2\pi/q_i$, so we consider
\begin{equation}\label{eq:pl_spectr}
S(q) = \left\{
                \begin{array}{lll}
                 0, &q>q_0\\
                 \mbox{const} \times q^{-(p+2)}, &q_i<q<q_0 \quad .\\
                  0, &q<q_i
                \end{array}
              \right.
\end{equation}

The parameters of the power-law spectrum with a Kolmogorov-like spectral index ($p=5/3$) can be determined from ground-based or in-situ observations \citep{1993JGR....98.1257T, 2001SSRv...97....9W, 2010MNRAS.402..362S, 2018ApJ...856...73C}. Similar to \cite{2007ApJ...671..894T} and \cite{2018ApJ...857...82K}, we assume that the density fluctuations form a power-law spectrum with a spectral index of $p=5/3$.

The inner scale of the inhomogeneities depends on the heliocentric distance as $l_i=r/(6.955\times 10^5)$ \citep{1987sowi.conf...55M, 1989ApJ...337.1023C} between $2-100R_\odot$, and the outer scale varies as $l_o=0.23\times r^{0.82}$ for distances $r$ from $4-80R_\odot$ \citep{2001SSRv...97....9W}.
Hence, the mean wavenumber can be written as a function of heliocentric distance as follows:
\begin{equation}\label{eq:q_bar1}
\overline{q\eps^2}\simeq 4\pi l_i^{-1/3} l_o^{-2/3}\eps^2=C_q r^{-0.88}\,,
\end{equation}
where $r$ is the distance from the solar centre in units of solar radius $R_\odot$, $\eps^2=\langle \delta n^2 \rangle/n^2$ is the variance of density fluctuations
and $C_q$ is a constant characterising the level of density fluctuations in units of $1/R_\odot$.
Thus, $C_q$, which characterises the density fluctuation properties, and $\alpha$, which characterises their anisotropy, are the main input parameters in our numerical models. 
By the definition, $\eps$ is the integral over all wave numbers while the density fluctuations have a broad spectrum which over all wave numbers is unknown. The $l_o$, and $l_i$, the outer and inner scales of the inhomogeneities are also hard to estimate.
Considering that the $\eps$, $l_o$, and $l_i$ are poorly known at the heliocentric distances of interest ($r=1.5-2 R_\odot$), we will use $C_q$ as a free parameter to match the observations.

\subsection{Ray-Tracing simulation}
\label{sec0202}
In our simulations, the radio emission source is assumed to be a point source at a heliocentric distance $r_0$. The source isotropically emits photons with a wavenumber $|\vec{k}|$, corresponding to the fundamental frequency $1.1 \omega _{pe}(r=r_F)$ or the harmonic frequency $2 \omega _{pe} (r=r_H)$, where $\omega_{pe}$ is the electron plasma frequency at the physical location of the source, i.e. at $r=r_F$ and $r=r_H$ for fundamental and harmonic frequencies, respectively. In this paper, we present simulations for emissions observed at 32.5~MHz, a frequency often imaged in LOFAR observations. The fundamental plasma emission of this frequency is generated at a heliocentric distance of $\sim$1.8 $R_{\odot}$, while the harmonic is generated at $\sim$2.2 $R_{\odot}$, according to the density model used (Equation (\ref{eq:density})). The background density in the upper corona is assumed to be
\begin{equation}
    \label{eq:density}
n(r)=4.8\times {10}^9 r^{-14} + 3\times {10}^8 r^{-6} + 1.39\times {10}^6 r^{-2.3}\,,
\end{equation}
which is an analytical approximation of the Parker density profile \citep{1960ApJ...132..821P}, with $r$ expressed in solar radii.

In each numerical experiment, $2\times 10^5$ rays are traced through the corona until all rays arrive at a sphere where the scattering is assumed to be negligible (i.e. $\omega _{pe} << \omega$). In our simulations, for a 32.5~MHz source the radius of the `scattering corona' is set to be $r_s\sim 9.6R_{\odot}$, where scattering becomes negligible, i.e. the total (photon path integrated) angular broadening between $r_s$ and the Earth is $0.'2$. This  allows us to calculate the typical observed angular size with a precision of $\sim 1\%$ for typical type III burst sizes of $20'$ at 32~MHz.
Therefore, the largest uncertainty in the results is the statistical error due to the finite number of photons in the simulations, which is accounted for and illustrated throughout our analysis.


In addition to scattering, we also take into account the free-free absorption of radio-waves according to \cite{1981phki.book.....L}. The observed brightness temperature is reduced by a factor of $e^{-\tau}$, where $\tau$ is the optical depth along the photon path in the simulations. The optical depth $\tau$ can be calculated as
\begin{equation}
    \label{eq:tau}
    \tau= \sum_{i=0}^{N}\sqrt{\frac{2}{\pi}}\frac{e^2 \mathrm{ln} \Lambda}{4m_ec{v_{Te}}^3\omega}\left(\frac{{\omega_{pe}}^4}{\sqrt{\omega^2-{\omega_{pe}}^2}}\right)\Delta s_i\, ,
\end{equation}
where $v_{Te}=\sqrt{k_BT_e/m_e}$ is the electron thermal velocity, $\Lambda$ is the Coulomb logarithm ($\sim$20 in the solar corona), and $\Delta s_i$ denotes the distance interval in each time step \citep[see][ for details]{2019ApJ...884..122K}.

\begin{figure}[ht!]
\includegraphics[width=\textwidth]{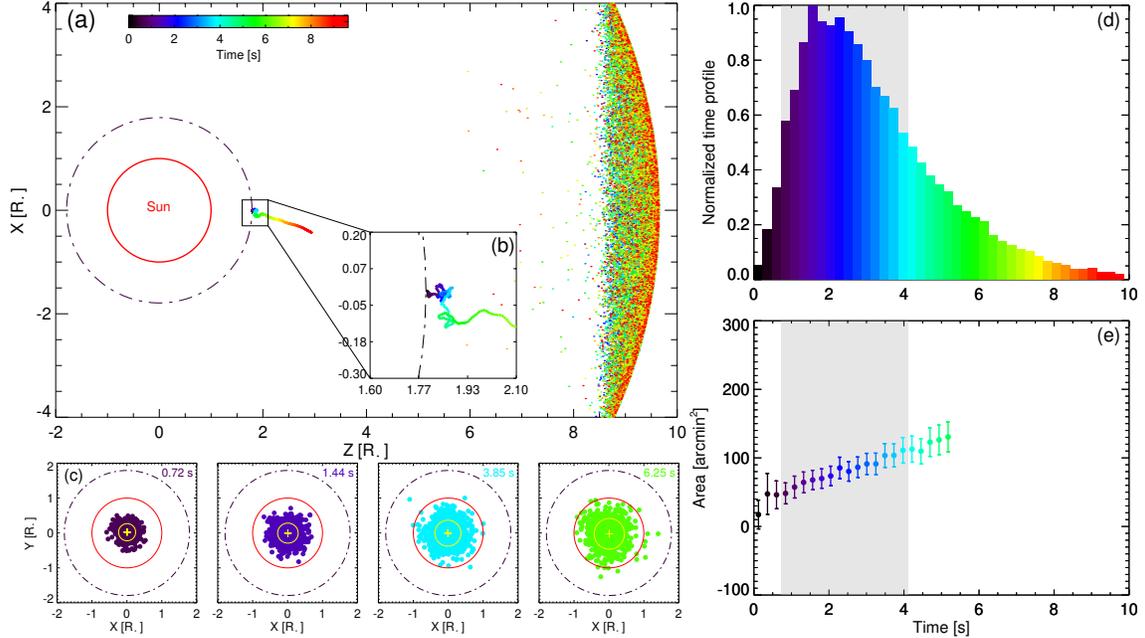}
\caption{Schematic diagrams illustrating the ray-tracing simulations conducted assuming isotropic scattering ($\alpha=1$) and turbulence levels given by $C_q=80R_\odot^{-1}$, showing a small number of photons for clarity.
The Sun is located at the center of the coordinate system with the Z-axis pointing towards the observer.
The dash-dotted circle illustrates the distance corresponding to the initial source location $r_0$. The source is located at $\theta =0^{\circ}$.
The different colors used represent different arrival times, where the free-space propagation time time has been subtracted.
Panel (a) shows a single-ray trajectory as a function of time after the photon's injection into the corona, and the photons arriving at distance $9.6R_{\odot}$ (where scattering becomes negligible).
Panel (b) is an enlargement of part of the single ray trajectory closer to the emission location.
Panel (c) shows snapshot images for the different time moments.
The yellow circle and plus sign represent the source's FWHM size and centroid location, respectively. The red circle represents the solar disk.
Panel (d) shows a histogram of the photon arrival times. The grey area indicates the FWHM of the impulse.
Panel (e) shows the FWHM size as a function of time. 
The error bars represent the one standard deviation uncertainty given by  Equations (47) and (48) in \citet{2019ApJ...884..122K}.\label{fig-1}}
\end{figure}

To illustrate the difference between anisotropic and isotropic ($\alpha$=1) turbulence, we perform simulations for  both.
The simulation results for $\alpha=1$ are shown in Figure~\ref{fig-1}. All simulated photons (observed at 32~MHz) are initially located at $r_0 \sim$1.8 $R_\odot$ for the fundamental plasma emission and $r_0 \sim$2.2 $R_\odot$ for the harmonic plasma emission.  Panel (a) shows the photon locations when they reach the boundary of the scattering corona, with different colors representing the time since the photon emission from the point source. The ray path of a randomly chosen photon is also illustrated (panel (b)).
The radio waves are strongly scattered close to the source, making wave propagation diffusive \citep[see e.g.][]{2019ApJ...873...33B}. The scattering rate decreases with distance, and the refraction due to large-scale density inhomogeneities becomes more significant, causing some `focusing' of the radio waves.

In order to derive the apparent location and shape of the source, all rays arriving at the boundary of the `scattering corona' are projected to the plane of sky to create an image (in this case, the plane perpendicular to the line-of-sight and containing the centre of the Sun). Photons with $0.85<k_z/k<1$ (at the moment of leaving the `scattering corona') are used to create the source intensity map $I(x,y)$ similar to \cite{2011A&A...536A..93J}. 
As done to observed emission images (see, e.g., \citep{2017NatCo...8.1515K}), the simulated radio images were fitted with a 2D Gaussian in order to determine the centroid positions and the full width at half maximum (FWHM) sizes, as shown in Figure \ref{fig-1}(c).

The number of photons arriving at the `scattering corona' ($r_s = 9.6 R_\odot$) varies with time as shown in Figure \ref{fig-1}(d), where the depicted time is offset by the free-space light prorogation ($(r_s -r_0)/c$), i.e., $t=t_{r_s}-(r_s -r_0)/c$. 
Thus, in the case of free-space photon propagation, all photons arrive at time $t=0$. However, due to the presence of density inhomogeneities giving rise to propagation effects, the observed pulse is broadened and the peak of the emission arrives later than free-space transport predicts. 
It should be noted that both the scattering and the large-scale refraction affect the photon propagation, but scattering produces the largest effect.
Figure \ref{fig-1}(d) demonstrates that for isotropic scattering with $C_q =80R_\odot^{-1}$, the radio pulse peak will be delayed by about 2.5~seconds and the instantaneously injected photons will be observed as a $\sim$3.5~sec long pulse. To show the dynamics of the radio source, images were simulated at different time moments, shown in Figure \ref{fig-1}(c).

\section{Comparison with LOFAR observations}
\label{sec03}

\subsection{Observations}\label{secobs}

Solar radio type III bursts are believed to be generated by nonthermal electrons propagating along open magnetic field lines. The ultrahigh temporal ($\sim$10 ms) and spectral (12.5 kHz) resolutions of LOFAR \citep{2013A&A...556A...2V} allow us to observe the unique fine structures (striae) in type III bursts emitted between 30-80 MHz. 
The type III burst presented in this study was observed on 2015 April 16 at $\sim$11:57 UT \citep{2017NatCo...8.1515K, 2018ApJ...856...73C, 2018ApJ...861...33K, 2018SoPh..293..115S}.  It is composed of hundreds of striae with a short duration ($\sim$1 s) and a narrow bandwidth ($\sim$100 kHz) for both its fundamental and harmonic branches. 
Background density perturbations are believed to have strong effects on the the formation of the striae \citep{2018ApJ...856...73C}. \cite{2017NatCo...8.1515K} illustrated that the radio-wave scattering effects dominate the observed spatial characteristics of the radio sources. To obtain a better understanding of the propagation of radio waves and the density inhomogeneity in the corona, we compare our radio-wave propagation simulations with the observed properties of the striated type III burst.

\subsection{Fundamental emission}\label{sec03F}

Simulations of emissions at 32.5~MHz reveal the time evolution, apparent motions and sizes of radio sources, which can be directly compared with the type III-IIIb LOFAR observations. To match the observations, we consider three parameters: the level of density fluctuations $C_q$, anisotropy $\alpha$, and the heliocentric angle $\theta$ of the source.

Similar to the observations, the temporal evolution of the sources at a given frequency is characterised in the decay phase of the temporal profile. We then apply exponential fits to the decay phase and describe the decay times as the half width at half maximum (HWHM) from the fits (similar to e.g. \citet{2018A&A...614A..69R, 2020ApJS..246...57K}). 
The one standard deviation uncertainty calculated during fitting is used as the error in the measurements of decay time.

\begin{figure}
\centering
\includegraphics[width=0.33\textwidth]{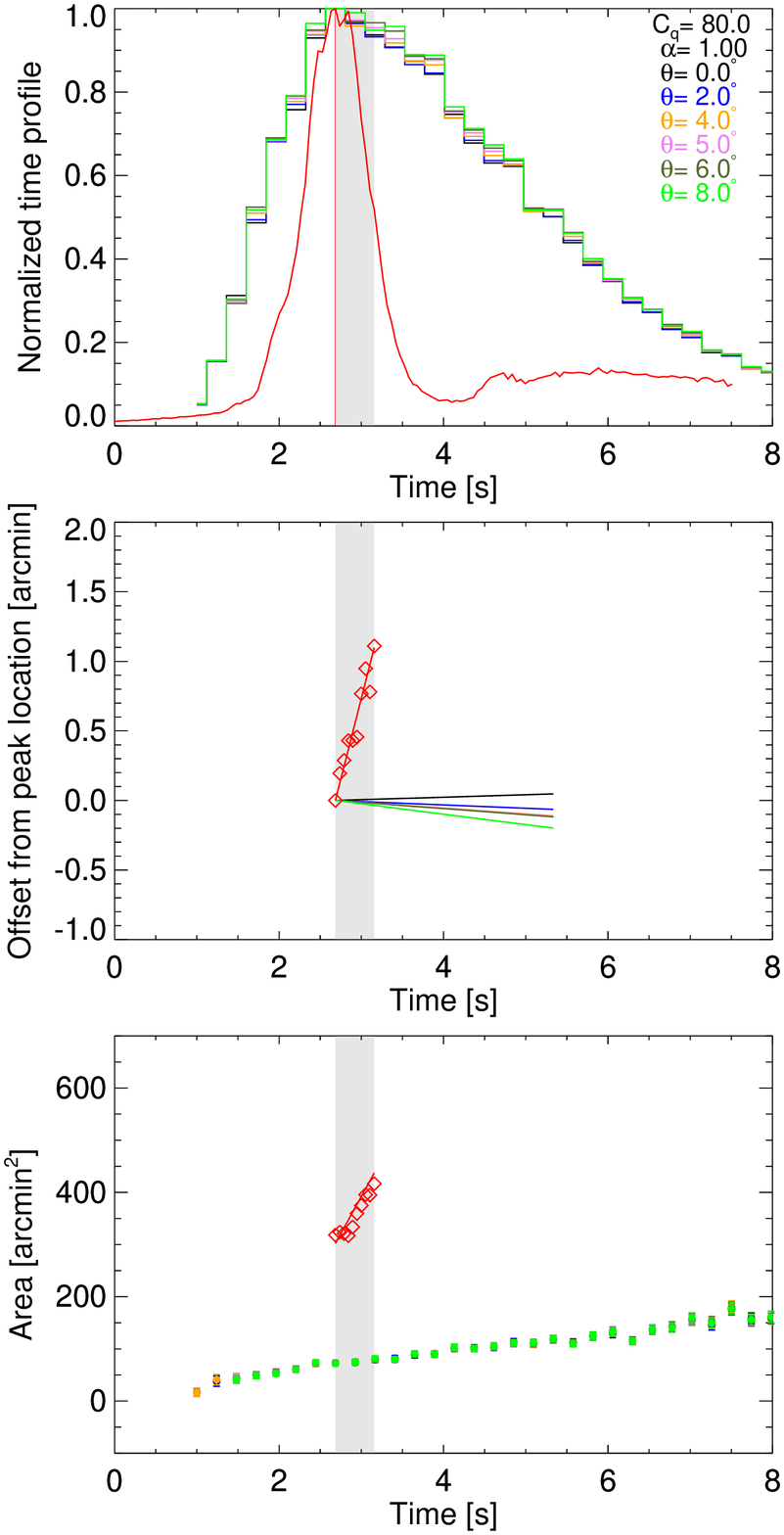}
\caption{Simulations of fundamental plasma emission from an instantaneous point source scattered by isotropic turbulence. The top panel shows the time profiles, the middle panel shows the centroid shifts, and the bottom panel shows the apparent source sizes with respect to the time after free-space propagation time. The frequency of the simulated source was taken to be 32.5 MHz, $C_q=80R_\odot^{-1}$, and the heliocentric angles $\theta$ of the source varied from $0^{\circ}$ to $8^{\circ}$, represented by the different colours, as indicated in the legend of the top panel. The red data represents LOFAR observations aligned to match the emission 
peak from \cite{2017NatCo...8.1515K}.
The grey-shaded area represents the decay time of the burst observed by LOFAR.
\label{fig-2}}
\end{figure}

Figure~\ref{fig-2} compares the main observed characteristics of type III sources with those produced by simulations for $\alpha=1$ (isotropic scattering) and $C_q=80$~$R_{\odot}^{-1}$. The grey-shaded area in all panels indicates the decay time derived from the observations. The top panels show the time profiles of the observed source (red curve) and of the simulated source with respect to the heliocentric angle. It can be seen that the FWHM of the impulse produced by the instantaneous source is broadened to about 3.5~s, which is substantially longer than the FWHM of the observed impulse (about 1.1~s). Similarly, the decay time in the simulations is approximately 2.5~s long, while the observed decay time is only $\sim$0.5~s.

The relative centroid positions were represented by the offset from the peak in order to better compare with Figure 4(b) from \cite{2017NatCo...8.1515K}.  From Figure \ref{fig-2}(b), the simulated source also demonstrates a much slower apparent motion compared to the observed source.
During the decay phase, the observed source moves by approximately 65~arcsec (in $\sim$0.5~s), while the simulated source---during the same period ($\sim$0.5~s)---moves by less than 5~arcsec, and moves only by 12~arcsec during the entirety of the simulated decay phase ($\sim$2.5~s).
Evidently, the source positions do not change substantially in the case of isotropic scattering, which is inconsistent with the LOFAR observations shown by the red line in Figure \ref{fig-2}.

The observed source areas, which were corrected for the FWHM area of the LOFAR beams ($\sim$ 110 $\mathrm{arcmin^2}$ at 32.5~MHz), vary from $\sim$300 to $\sim$440 $\mathrm{arcmin^2}$ during the decay phase ($\sim 0.5$s for the fundamental emissions). From the isotropic scattering simulations,
the simulated size for the fundamental emission ranged
from $\sim$60 to $\sim$100~$\mathrm{arcmin^2}$ during the 2.5~s of the decay phase, which is substantially different from the LOFAR observations.

Although the simulated decay time was longer than the observed, 
the apparent source size was four times smaller. Therefore, under the isotropic scattering assumption, the simulated decay time and source size cannot both agree with the observations, irrespective of how weak or strong the scattering effects are set to be, given that stronger scattering will result in both larger source sizes and longer decay times. Furthermore, in the case of isotropic scattering, the simulated apparent motion of centroids was much smaller than that observed. 
Therefore, it is clear that isotropic scattering cannot simultaneously explain the size and temporal evolution of a typical type III radio source (as illustrated by \cite{2019ApJ...884..122K}) and should not be used for simulations of radio-wave transport.

\subsection{Anisotropic scattering}

\begin{figure}[ht!]
\includegraphics[width=0.99\textwidth]{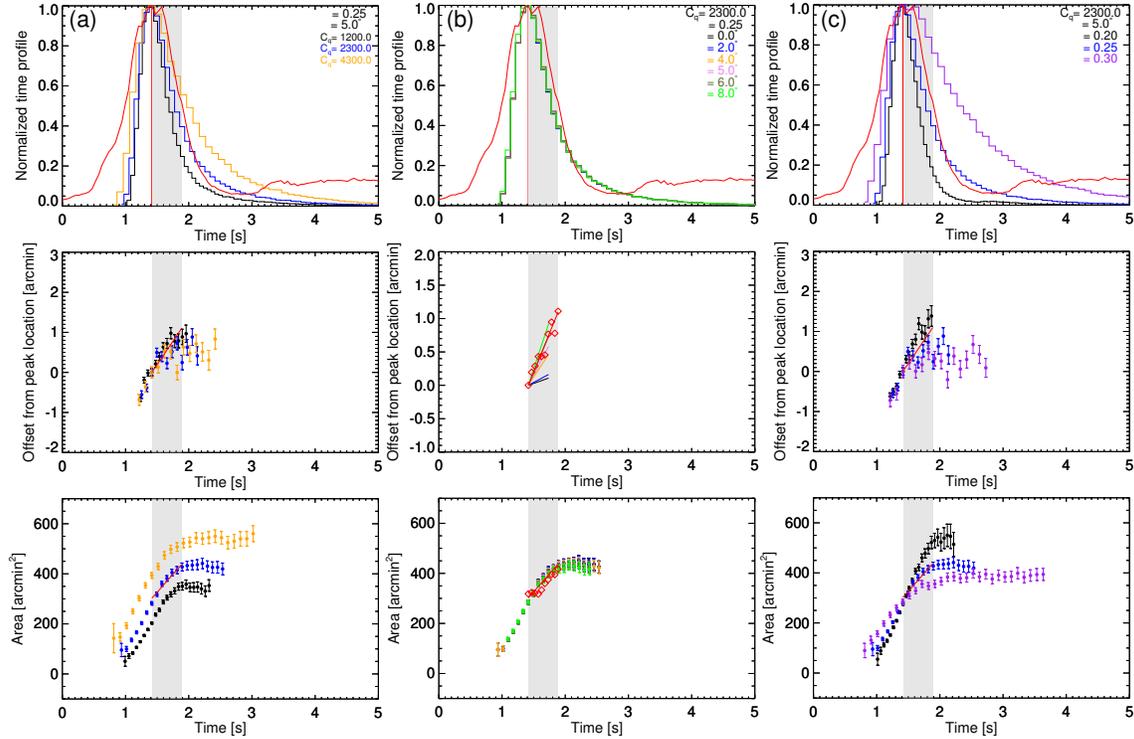}
\caption{Simulations of fundamental plasma emission from an instantaneous point source scattered by anisotropic turbulence. The top row shows time profiles, the middle row shows the shifts of the apparent source centroids, and the bottom row shows apparent source areas with respect to time (where the free-space prorogation time has been subtracted).
Each column shows results for different sets of parameters shown in the top panels.  Column (a) shows the effects of  different mean wavenumbers of the density fluctuations for $C_q=$~1200, 2300, and 4300~$R_{\odot}^{-1}$, column (b) shows the effect of varying the heliocentric angle $\theta$ of the source from $0^{\circ}$ to $8^{\circ}$, and column (c) shows the effect of varying the anisotropy from $\alpha$=0.2 to 0.3.
The red lines and symbols represent measurements made from the LOFAR observation of the type III-IIIb radio burst at 32.5 MHz, and the grey-shaded area represents the observed decay time.
The error bars indicate the one standard deviation uncertainties.\label{fig-3}}
\end{figure}

Given that the isotropic description of scattering was shown to be insufficient, we simulate anisotropic scattering with anisotropy values $\alpha$ ranging from 0.2 to 0.3, so that the perpendicular density fluctuations have a stronger effect on the radio-wave propagation than the respective parallel component 
(see Equation \ref{eq:S_ani} and \cite{2019ApJ...884..122K}). 
At the same time, we investigate different relative density fluctuation levels for fundamental emissions, giving spectrum-averaged mean wavenumbers $C_q=$~1200, 2300, and 4300~$R_{\odot}^{-1}$ from Equation \ref{eq:q_bar1}.
The temporal evolution, apparent sizes, and locations of radio sources simulated assuming an anisotropic turbulence with an initially instantaneous point-like emission 
are shown in Figure \ref{fig-3}.  
In the top panels, the thin red line represents the temporal evolution of the observed type III-IIIb radio source, with the grey-shaded area indicating its decay phase.

The decay times are $\sim$0.32, 0.50, and 0.72 for the three used values of $C_q$, respectively. Higher levels of density fluctuations result in longer decay times and larger FWHM sources sizes. For the case of $C_q=2300R_{\odot}^{-1}$, the source size changes from $\sim$ 280 to $\sim$ 430 in 0.5 seconds during the decay phase (blue-colored dots at the bottom of Figure \ref{fig-3} (a)), which agree with the values obtained from the LOFAR observations shown by the red line in Figure~\ref{fig-3} \citep{2017NatCo...8.1515K,2018SoPh..293..115S}.

The heliocentric angle $\theta$ affects the apparent centroid position but has almost no influence on the time profiles and source sizes, as seen in Figure \ref{fig-3}(b). The centroid shifts (relative to the centroid positions at the peak times) are shown in the middle row, with the red line corresponding to the LOFAR observations.

The effects of the anisotropy parameter $\alpha$ (defined in Equation \ref{eq:A_matrix}) on the observed source properties are demonstrated in Figure \ref{fig-3}(c) for $\alpha=0.2, 0.25$, and $0.3$. The decay times are 0.24, 0.50, and 1.01 s, respectively,  and the apparent source sizes range from $\sim$ 270 to 400 $\mathrm{arcmin^2}$, $\sim$~280 to 430~$\mathrm{arcmin^2}$, and $\sim$ 280 to 380 $\mathrm{arcmin^2}$ during the decay phase, respectively, for the different anisotropy values. The results show that stronger anisotropy yields a shorter time duration and a larger areal expansion rate.

The best fit for the fundamental source observed at 32~MHz assuming instantaneous emission from a point source is provided by the model in which the point source is located at a heliocentric angle $\theta= 5^{\circ}$, turbulence with anisotropy $\alpha=0.25$, and $C_q=2300$~$R_{\odot}^{-1}$. The comparison of the observed and simulated lightcurves suggests that the intrinsic duration of fundamental emission cannot be longer than $0.3$~s, otherwise the observed profile will be too long. For gaussian sources, the observed source area is the sum of the intrinsic area and the area due to scattering.  A comparison of the observations to the simulations reveals that intrinsic areas smaller than $\sim 50$~arcmin$^2$; larger intrinsic sizes would contradict observations producing the expansion rate below the observed rate.

\subsection{Harmonic emission}

\begin{figure}[ht!]
\includegraphics[width=0.99\linewidth]{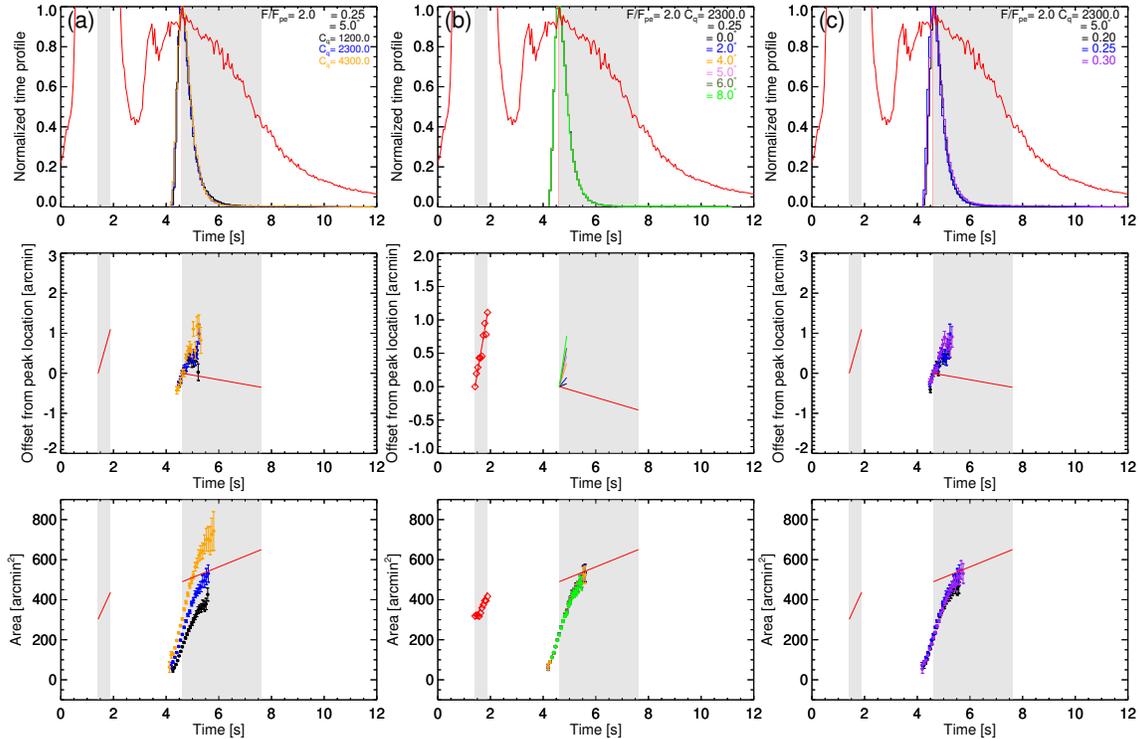}
\caption{Same as Figure \ref{fig-3}, but for an instantaneous point source of harmonic emission observed at 32.5 MHz. The simulation parameters are the same as the parameters in Figure \ref{fig-3}. \label{fig-4}}
\end{figure}

We also simulate the harmonic component of radio waves emitted 
at $2.2R_{\odot}$ (where the local plasma frequency is $16$~MHz) propagating in an anisotropic turbulence medium, and we compare the results with the harmonic source observed by LOFAR at 32.5~MHz. Figure~\ref{fig-4} shows the simulation results obtained by varying $C_q$ (column (a)), the heliocentric angle (column (b)), and the level of anisotropy (column (c)). The parameters used for Figure \ref{fig-4}  are the same as the parameters used when simulating the fundamental emission component (see Figure \ref{fig-3}).

Simulations with $C_q=2300R_\odot$, $\alpha =0.25$, $\theta =5^o$, and an instantaneous point source produce a time profile, source positions, and areas which are similar to the results of the fundamental emission simulations.Due to propagation effects, harmonic emission is found to have a $\sim$0.4 s decay time, which is much shorter than the 3~s observed by LOFAR  (red thin line in Figure \ref{fig-4}).
The simulated area of the harmonic emission source is about $\sim 300$~arcmin$^2$ near the peak, which is comparable to the fundamental emission source area, but smaller than the observed $\sim 500$~arcmin$^2$.
Furthermore, the simulated harmonic source area changes rapidly with time, and the centroids demonstrate rapid motion, similar to that of the fundamental emission sources but inconsistent with the LOFAR observations of the harmonic component.

All these results (the discrepancy in the time profiles, source motions, and sizes) suggest that the harmonic emission cannot be explained by an instantaneous point source and that it has a finite time duration and size.
Indeed, if the fundamental emission is produced at $1.8R_\odot$ 
and the harmonic at $2.2R_\odot$, i.e. $\Delta r= 0.4R_\odot$ away from the fundamental source, the time-of-flight spread of electrons 
could produce a finite harmonic emission duration. 
According to the observations, the drift rate of the type IIIb burst gives an electron speed 
of approximately $c/3$, where $c$ is the speed of light \cite[see][for details]{2017NatCo...8.1515K}. 
If the electron beam has a uniform spread of velocities between $c/6$ and $c/3$, 
the time-of-flight duration of electrons at the harmonic location would be $\Delta t=  3\Delta r/c \sim3$~s. Self-consistent simulations of electron transport \citep{2018A&A...614A..69R} support such an expansion of the electron beam and an increase of the emission duration with distance.
In addition, there is a finite time required for the production 
of harmonic emission in a given location \citep[e.g.,][]{2014A&A...572A.111R,2018PhPl...25a1603Y}.
Therefore, when the estimated electron time-of-flight is taken into account, the duration of harmonic emission could be 3-4~s.

\begin{figure}
\centering
\includegraphics[width=0.5\textwidth]{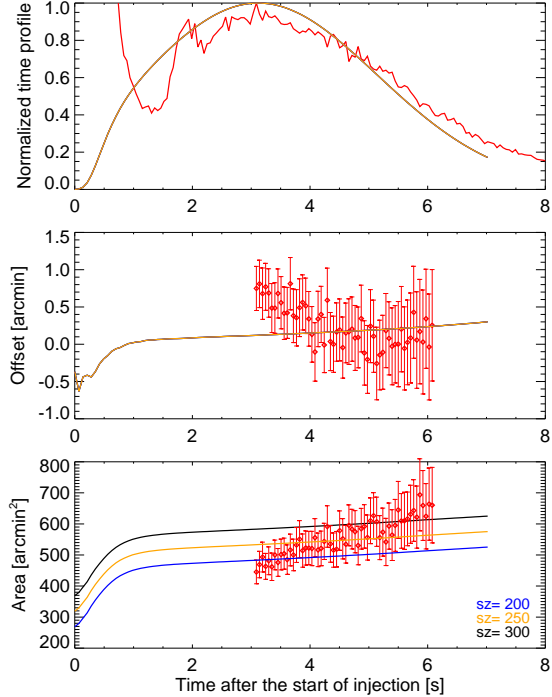}
\caption{From top to bottom, the solid black lines show the time profile, the centroid position, and area for the harmonic emission obtained assuming a Gaussian profile
$\exp(-(t-t_0)^2/(2\times 2^2))$
and a finite area of 200, 250, $300$~arcmin$^2$.
All other parameters are the same as those used in Figure \ref{fig-4}.
The red data represent the observed time profile, centroid position, and size for the harmonic component of the type III-IIIb burst observed with LOFAR, and the associated errors.
\label{fig-5}}
\end{figure}

To simulate the effect of finite harmonic emission time, we assume that the harmonic emission is a Gaussian pulse $\exp(-t^2/2\sigma^2)$ with $\sigma =2$~sec (see the caption of figure \ref{fig-5}). Then the observed profile is the convolution of the intrinsic emission and the broadening due to scattering.
In addition, we assume that the harmonic source has a finite emission area of 200, 250 and $300$~arcmin$^2$.
The results are shown in Figure~\ref{fig-5}. 
Importantly, it can be seen that prolonged emission at the source results in a smaller motion of the centroid compared to the instantaneous injection in Figure~\ref{fig-4}.
Moreover, while the instantaneous harmonic source shows a fast source motion (Figure \ref{fig-4}), the 4-second harmonic emission does not have a clear motion, which is more consistent with the LOFAR observations.

The comparison of the observed harmonic emission with the simulations suggests that an emission source with the physical area of up to $\sim 200$~arcmin$^2$ at the peak time will be consistent with the observations. 
At the same time, in order to explain the slow centroid motion of harmonic emissions and the areal expansion, a continuous harmonic emission lasting $\sim$4 seconds is required.

\section{Summary}
\label{sec04}

We quantitatively investigated the way in which scattering of radio-waves on random density fluctuations with a power-law spectrum affects the time profile evolution, sizes, and positions of the observed radio bursts emitted via the plasma emission mechanism. Although comparison was made to type III bursts, 
similar arguments are applicable to all bursts emitted via the plasma emission mechanism, e.g. noise storms \citep[e.g.][]{2015A&A...576A.136M}, 
type II solar radio bursts \citep[e.g.][]{2020ApJ...893..115C}, 
drifting pairs \citep[e.g.][]{2019A&A...631L...7K, 2020ApJ...898...94K}, and probably type IV bursts \citep[e.g.][]{2019ApJ...873...48G}.

Density fluctuations in the corona lead to angular broadening of the sources, so that the observed source area is increased by scattering. 
Larger density fluctuations (i.e. larger $C_q$ values) would also lead to a longer duration of the wave propagation and increase the duration of the observed emission pulse. As a result, the observed decay times and the observed peak time of the emission increase with higher levels of density fluctuations.

Unlike isotropic turbulence, anisotropic turbulence produces an apparent motion of the source with time, with the apparent velocity depending on the source heliocentric angle (i.e. the angle between the line-of-sight and the direction from the centre of the Sun to the physical source position). 
Sources with large projection angles (i.e. located close to the limb) experience larger radial displacements, as shown in Figure \ref{fig-3}. 
The time profile remains rather similar for the small heliocentric angles studied. Strong anisotropy ($\alpha \ll 1$, i.e. density fluctuations being mostly perpendicular) means that the effects occur predominantly along the perpendicular direction to the radial magnetic field. This effect is similar to the observed
ellipticity of galactic radio sources observed through the solar corona, 
where sources are broadened more strongly along the tangential to the solar limb direction \citep[e.g.][]{1958MNRAS.118..534H,1972PASAu...2...86D}.

\begin{figure}
\centering
\includegraphics[width=0.7\textwidth]{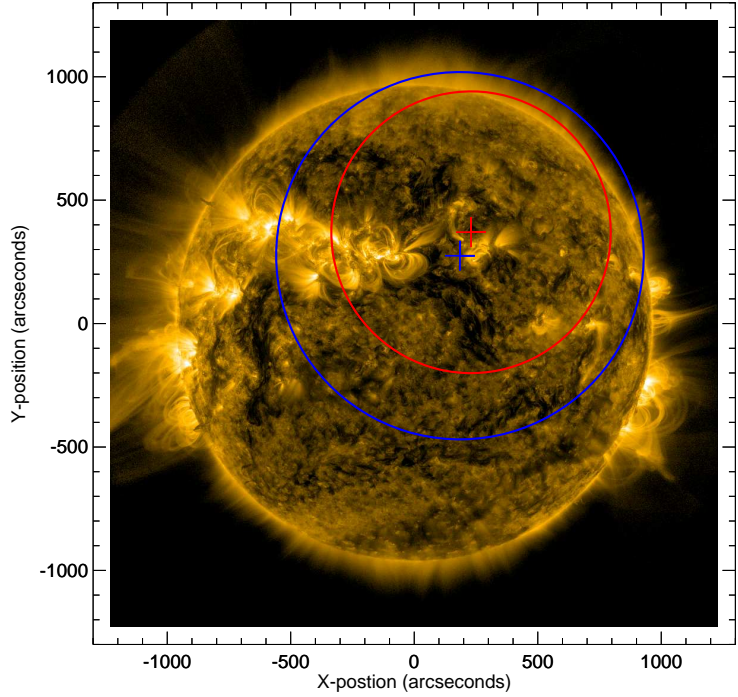}
\caption{
Simulated fundamental (red contour) and harmonic (blue contour) radio sources, overlaid on an SDO/AIA 171\AA\ image showing the solar surface (at $\sim$11:57~UT) in EUV wavelenghts.
The sources were simulated using $C_q \sim 2300 R_{\odot}^{-1}$, $\theta = 5^{\circ}$, and $\alpha \sim$ 0.25.  The harmonic source was simulated using the injection profile of a finite emission area of 200~arcmin$^2$ shown in Figure \ref{fig-5}.
\label{fig-6}}
\end{figure}

By matching the main characteristics (size, position, and time profile) of the observed sources with those observed by LOFAR, we have estimated the properties of plasma turbulence in the solar corona. The simulations reproduced the sub-second evolution of the source area, position, and time profile of both fundamental and harmonic radio sources. We find that radio-wave scattering due to turbulence with spectrum-weighted mean wavenumbers of density fluctuations $C_q \sim 2300 R_{\odot}^{-1}$, a source heliocentric angle $\theta = 5^{\circ}$, and an anisotropy $\alpha \sim$ 0.25, can explain the decay time, apparent source motion, and the temporal evolution of the source size for the fundamental emission of the type III-IIIb radio burst observed by LOFAR.  
We also simulated the harmonic emission using the same parameters as for the fundamental emission, and found that it can be explained by the same parameters of turbulence when the harmonic has a finite source and finite emission time. The intrinsic source, which is located at a heliocentric angle $\theta = 5^{\circ}$, has an area of $200$~arcmin$^2$ at the peak time (i.e. FWHM diameter $\sim 16'$ ) without scattering and has a FWHM duration of 4.7 seconds.
Then the observed size due to scattering becomes $480$~arcmin$^2$ 
with a slightly longer duration of 4.8~sec. Another interesting result is that a continuous emission of harmonic radiation for over $4$~s is required to explain
the small change in position and size of the harmonic source at 32.5~MHz.
The intrinsic duration of harmonic emission is likely to be related to electron transport effects.
Using the obtained parameters, we produced an image showing the fundamental and harmonic sources (imaged at 32.5~MHz), presented in Figure \ref{fig-6}.  It is evident that the simulated sources successfully reproduce the observed sources, shown by Figure 2 in \cite{2017NatCo...8.1515K}.

The agreement with the LOFAR observations suggests that $C_q=2300$~$R_{\odot}^{-1}$, so when the density fluctuations are $\eps\simeq 0.1$ \citep[e.g.][]{2018ApJ...860...34H},  the characteristic density scale $h$ should be about $5.5\times 10^{-5}R_\odot\sim 38$~km.
In the present work, we used an extrapolation of the inner and outer scale measurements from distances $>2R_\odot$, but a more rigorous analysis and multi-frequency observations are needed to determine the spatial behaviour of $\overline{q\eps ^2}$ throughout the corona and the solar wind.

The results and conclusions of our study can be summarised as follows:
\begin{enumerate}[label=\roman*.]
    \item Isotropic scattering cannot simultaneously describe all the observed radio emission characteristics, but instead an anisotropic scattering description is required to do so.
    \item The apparent source displacement becomes larger for larger heliocentric angles $\theta$, whereas the source area and the time profile are affected to a lesser extend.
    \item The time evolution of the radio flux, source size, and centroid motions of a type III-IIIb burst, for both its Fundamental and Harmonic components, can be successfully reproduced using the ray-tracing simulations applied.
    \item The spectrum-weighted mean wavenumber of the density fluctuations ($C_q$), the level of anisotropy ($\alpha$), and the heliocentric angle ($\theta$) of a type III-IIIb burst are estimated, by comparing its observed characteristics to radio-wave propagation simulations.
\end{enumerate}
\noindent
The comparison of simulations to observations of source sizes and positions at sub-second time scales allows us to understand the propagation of radio waves in the corona. Radio-wave propagation effects should be taken into account when inferring physical parameters using solar radio-imaging observations, given that the observed radio properties do not represent the intrinsic source properties. The simulations presented are applicable for all types of radio bursts that are generated via the plasma emission mechanism. We note, however, that the density turbulence may not be same for different events and different frequencies.  Therefore, a statistical study of spectral and imaging radio observations that are compared with radio-wave propagation simulations is required in the future.

\acknowledgments
The work is supported by NSFC grants 11790301, 11973057, 12003048 and National Key R\&D Program of China 2018YFA0404602.


\bibliography{paper}
\end{document}